# The Influence of the Second Harmonic in the Current-Phase Relation on the Voltage-Current Characteristic of high $T_C$ DC SQUID

**Ya. S. Greenberg[1,*], I. L. Novikov[1], V. Schultze[2], H.-G. Meyer[2]**

[1]Novosibirsk State Technical University, 20 K. Marx Ave. Novosibirsk, 630092, Russia

[2]Institute for Physical High Technology, D-07702, Jena, Germany

A theory for the voltage-current characteristic in high $T_C$ DC SQUIDs, which accounts for a second harmonic in the junction current-phase relation, is developed. The comparison with experiment is performed. It is shown that if the second harmonic is prevailed, the theory can explain the large deviations of the experimental voltage modulation from theoretical predictions and computer simulations based on conventional sinusoidal current-phase relation.

PACS: 74.50.+r; 85.25.-j; 85.25.Dq

## 1 Introduction

As is well known, there exists a significant discrepancy between experimental results and numerical simulations of the voltage-to-flux transfer function of high $T_C$ DC SQUIDs [1,2]. This is one of the most important unsolved problems, which seriously hinders the optimization of high $T_C$ DC SQUIDs for applications.

Inspite of extensive computer simulations [1, 3] and theoretical studies [4, 5, 6] that have been performed in the attempt to predict reliably transfer function and energy resolution of high $T_C$ DC SQUIDs, a marked disagreement with experiment still exists: experimental transfer functions in many cases are much lower than the values predicted by theory and computer simulations; the white noise is about ten times higher than predicted.

One of the possible reasons for these discrepancies could be attributed to the junction asymmetry of SQUID interferometer (unequal critical currents or (and) normal resistances ), which for grain boundary junctions is about 20%−30% due to on chip technological heterogeneity. However, the

---

*Corresponding author, e-mail: ya_greenberg@lycos.com; after 30 August: greenbergy@online.nsk.su.

junction asymmetry can explain only small deviations from theoretical curves [3, 7]. Therefore, the problem of large deviation of transfer function and the voltage modulation relative to theoretical predictions is still open.

The other reason for aforementioned deviations, which is investigated in the paper, is the influence of a second harmonic in the current-phase relation (CPR) of a high temperature superconducting Josephson junction on the voltage modulation in high $T_C$ DC SQUID.

It is well established now that the CPR in high $T_C$ junctions is non-sinusoidal, possessing a second harmonic component whose relative magnitude depends on mutual orientation of the d-wave superconductors, and whose sign can be temperature dependent [8, 9, 10]:

$$I_S = I_1 \sin \varphi + I_2 \sin 2\varphi \quad , \tag{1}$$

where $I_S$ is a supercurrent flowing through the junction, $\varphi$ is a phase difference of the superconducting order parameter across the junction. The amplitude of the first harmonic $I_1$ depends on the relative orientation of d-wave superconductors. In asymmetric 45° [001]-tilt grain boundary junction the amplitude $I_1$ is expected to disappear due to symmetry [11]. In this case the amplitude $I_2$ becomes prominent, and has been observed in direct CPR measurements [12, 13], and as a half-flux quantum periodicity of critical current in YBCO DC SQUIDs [14].

More recent studies revealed a substantial amount of a second harmonic also in [001]-tilt grain boundary junction with a 30° asymmetric misorientation angle [15].

In the present paper we investigate the effect of a second harmonic on the voltage-current and voltage-flux curves of high $T_C$ DC SQUID.

We consider a symmetrical DC SQUID interferometer with equal shunt resistances, $R_1=R_2=R$, and with equal amplitudes of each harmonics:

$$I_1^{(1)} = I_1^{(2)} \equiv I_1; \quad I_2^{(1)} = I_2^{(2)} \equiv I_2 \quad , \tag{2}$$

where the superscripts refer to the junction number.

## 2 DC SQUID with small inductance

First detail experimental investigation of the influence of second harmonic on the dynamic behavior of small inductance YBCO DC SQUID with asymmetric 45° grain boundary junctions was performed in [16]. It was shown that the peculiarities in the dependence of critical current on the external flux can be explained by the large amount of second harmonic in the junction CPR.

In this section we obtain the analytical expression for the output voltage of small inductance DC SQUID, which allows one to investigate the SQUID properties in a broad range of temperatures, and of the amplitudes $I_1$ and $I_2$.

Neglecting the inductance of the loop we can write the energy of DC SQUID loop in the external magnetic field:

$$U = -E_J \left( ij + 2\cos\varphi_X \cos\varphi + g \cos 2\varphi_X \cos 2\varphi \right), \quad (3)$$

where $E_J = \Phi_0 I_1/2\pi$ is the Josephson coupling energy, $i=I/I_1$, $I$ is the bias current, $\varphi_X=\pi\Phi_X/\Phi_0$, $\Phi_X$ is the external magnetic flux, $\Phi_0=h/2e$ is a flux quantum, $\gamma=I_2/I_1$.

As is known, DC SQUID with a negligible small inductance can be considered as a single Josephson junction with normal resistance $R/2$ and doubled critical current. Therefore, we may use the known approach of Ambegaokar and Halperin [17] to account for a second harmonic in the voltage current characteristic of such a SQUID at finite temperature:

$$\frac{V}{RI_1} = \frac{\pi \Gamma}{p(i, \Gamma, \varphi_X)} \quad (4)$$

where $\Gamma = 2\pi k_B T/\Phi_0 I_1$ is the noise parameter, T is the absolute temperature,

$$p(i,\Gamma,\varphi_X) = \left[ \int_0^{2\pi} e^{-W(y)} dy \int_0^y e^{W(x)} dx - \frac{1}{1-e^{\frac{2\pi i}{\Gamma}}} \int_0^{2\pi} e^{W(x)} dx \int_0^{2\pi} e^{-W(x)} dx \right] \quad (5)$$

$$W(x) = (i/\Gamma)x + (2/\Gamma)\cos\varphi_X \cos x + (g/\Gamma)\cos 2\varphi_X \cos 2x \quad (6)$$

The expression (4) is valid for any temperature below $T_C$, the critical temperature of a superconductor, however, all calculations we show below have been performed for T=77 K.

The equation (4) is a good approximation for DC SQUIDs with $\alpha \ll 1$, $\beta \ll 1$, where $\alpha = L/L_F$, $L_F = (\Phi_0/2\pi)^2/k_B T$ is a fluctuation inductance which is equal to 100 pH at T=77 K, $\beta = 2LI_1/\Phi_0$. Therefore, in general, SQUID behaviour is described by three parameters, $\alpha$, $\beta$, $\Gamma$, but only two are independent due to relation $\alpha = \pi\beta\Gamma$.

The voltage-flux curves (VFC) calculated from the expression (4) are shown on Fig. 1. As is seen, the presence of a second harmonic gives rise to the half flux quantum periodicity of VFC. It also implies the voltage invariance under reversal of the sign of $I_2$. In addition, the second harmonic reduces the modulation depth.

**3 DC SQUID with large inductance**

As is well known the high $T_C$ DC SQUIDs, which are used in practice for measuring devices, have large geometrical inductance of interferometer loop, typically L>100 pH.

Therefore, in this section we consider a DC SQUID with large inductance, $\alpha \geq 1$ and any $\beta$ and $\Gamma$ consistent with the relation $\alpha = \pi\beta\Gamma$. The theory of the voltage-current characteristics for such SQUID has been developed earlier [5]. Here we apply this theory to a SQUID with the second harmonic in junctions CPR. According to the approach of [5] we have obtained the following result for the voltage across a SQUID with a second harmonic in CPR:

$$\frac{V}{RI_1} = J - \exp(-a/2)\cos(2j_X)f(i,\Gamma,g) \qquad (8)$$

where

$$J^{-1} = 2i\sum_{n=-\infty}^{n=+\infty} \frac{(-1)^n G_n^+ G_n^-}{i^2 + 4n^2\Gamma^2} \qquad (9)$$

$$f(i,\Gamma,g) = 8J^3\left[i^2 A(A+C) - 4\Gamma^2 B(D-B)\right] \qquad (10)$$

$$A = \sum_{n=-\infty}^{n=+\infty} \frac{(-1)^n G_{n+1}^+ G_n^-}{i^2 + 4n^2\Gamma^2} \qquad (11a)$$

$$C = \sum_{n=-\infty}^{n=+\infty} \frac{(-1)^n G_n^+ G_{n+1}^-}{i^2 + 4n^2\Gamma^2} \qquad (11b)$$

$$B = \sum_{n=-\infty}^{n=+\infty} \frac{(-1)^n n G_{n+1}^+ G_n^-}{i^2 + 4n^2\Gamma^2} \qquad (11c)$$

$$D = \sum_{n=-\infty}^{n=+\infty} \frac{(-1)^n n G_n^+ G_{n+1}^-}{i^2 + 4n^2\Gamma^2} \qquad (11d)$$

$$G_n^\pm = \sum_{m=-\infty}^{m=+\infty} (\pm 1)^m I_{n-2m}(1/\Gamma) I_m(g/\Gamma) , \qquad (12)$$

where $I_n$ is a modified Bessel function.

The voltage modulation $\Delta V = V(\varphi_X=\pi/2) - V(\varphi_X=0)$ is readily obtained from (8):

$$\frac{\Delta V}{RI_1} = 2\exp(-a/2)f(i,\Gamma,g) \qquad (13)$$

The expression (8) is valid for α≥1 and any values of β and Γ, which are consistent with the condition α=πβΓ. Therefore, it can be applied for the analysis of a majority of practical high $T_C$ DC SQUIDs with Γ≈ 0.05 – 1, β≥ 1, α≥ 1. However, it should be remembered that (8) is the approximate expression which accounts for the first order term in the perturbation expansion of the voltage over small parameter ε = exp(−α/2) (see [5]).

Contrary to the case of zero inductance (expression (4)), here VCC (8) is not invariant under sign reversal of $I_2$ since magnetic energy term in the Hamiltonian of DC SQUID, ($\Phi-\Phi_X$)²/2L, destroys the invariance. Indeed, as follows from (12) $G_n^{\pm}(-g) = G_n^{\mp}(g)$, then, the quantities in (8) transforms as follows: $J \to J, A \to C, C \to A, B \to D, D \to B$. Therefore, $f(i,\Gamma,-g) \neq f(i,\Gamma,g)$. However, for large inductance ($\alpha \geq 1$) the VCC is approximately invariant under sign reversal of $I_2$ since the second term in right hand side of (8) is much smaller than the first one.

If the second harmonic is absent ($\gamma=0$), then, we have $G_n^{\pm} = I_n(1/\Gamma)$, C=A, D=B, and we get the result obtained in [5] for symmetric DC SQUID with conventional CPR.

It is interesting to note that if the first harmonic is absent ($I_1=0$), then the voltage modulation is, within our approximation, exactly equal to zero. It follows from (12) that for $I_1=0$

$$G_n^{\pm} = \begin{cases} (\pm 1)^{n/2} I_{n/2}(g/\Gamma) & \text{for } n \text{ even} \\ 0 & \text{for } n \text{ odd} \end{cases}, \qquad (14)$$

Then, from (11) we have *A=C=D=B=0*. Therefore, the modulation signal (the second term in (8)) is absent in this case. Of course, it is not a strong statement, since the expression (8) accounts only for the first order term in the perturbation expansion of the voltage over small parameter $\varepsilon = \exp(-\alpha/2)$. However, from these considerations we can expect a significant reduction of the voltage modulation if the second harmonic is prevailed.

The influence of the second harmonic on the voltage-current curve (VCC) is shown in Fig. 2. We see that the second harmonic enhances the SQUID critical current. The more the amplitude of the second harmonic the more is the enhancement of the critical current. As is seen from Fig. 2, for given value of the sum, $I_1+I_2$, the critical current is greatly enhanced, if the second harmonic is prevailed.

As is seen from (13) the influence of inductance on the voltage modulation is factored out, so that below we consider the reduced modulation $\exp(\alpha/2)\Delta V/R$, which depends on the bias current I and harmonic amplitudes $I_1$ and $I_2$.

In the absence of the second harmonic ($I_2=0$) the reduced voltage modulation vs-bias current curves, $\Delta V(I)$ have the form as shown in Fig. 3. For a given junction critical current $I_1$ the curve has a well defined maximum $\Delta V(I_{MAX})=\Delta V_{MAX}$ at the corresponding value of the bias current I=$I_{MAX}$.

Our calculations show that the admixture of a second harmonic independently of its sign reduces the voltage modulation as is seen from Fig. 4. A significant reduction is obtained only if the amplitude of the second harmonic is several times that of the first harmonic. However, the forms of the curves are similar to those for $\gamma=0$: on the bias current axes they have one maximum, which position is shifted to lower bias currents for relatively small admixture of the second harmonic, and to the higher bias currents if the second harmonic is strongly prevailed.

**4 Comparison with experiment**

For the practical purpose it is convenient to have a set of the values of maximum voltage modulation $\Delta V_{MAX}$ with a corresponding values of the bias current $I_{MAX}$. Just these two quantities can easily be measured in practice: by varying the bias current the point of maximum voltage modulation $\Delta V_{MAX}$ is being obtained at particular value of bias current $I=I_{MAX}$.

From equation (13) for a set of two fixed values, $I_1$ and $\gamma$ we computed a set of the values of maximum voltage modulation $\Delta V_{MAX}$ with a corresponding values of bias current $I_{MAX}$. In this way we obtained the table of values $\Delta V_{MAX}(I_1, \gamma, I_{MAX})$. With the aid of the table we draw for several $\gamma$-s the dependences $\Delta V_{MAX}(I_{MAX})$, which are shown in Figs. 5. The different points on a particular curve belong to different values of $I_1$. As it follows from Fig. 5, a relative small portion of the second harmonic gives rise to a relative small reduction of the voltage modulation as compared to that for the conventional CPR. A substantial reduction is obtained only if the second harmonic is strongly prevailed.

In order to compare our theory with experiment we took the same 70 DC SQUIDs with $\alpha \geq 1$ and $\Gamma \geq 0.05$, which have been chosen before for the same purpose [7]. Most of these SQUIDs are single layer ones, using 100 or 200 nm thick $YBa_2Cu_3O_{7-x}$ films deposited by laser ablation onto $SrTiO_3$ bicrystal substrates with 24° and 30° misorientation angles. The technology is described in detail in [18]. The single layer SQUID layouts cover small SQUIDs either used solo [19] or coupled to a pickup loop. Direct as well as inductive coupling were used, with pickup loops occupying a quarter of the complete 10 mm x 10 mm substrate, respectively [20].
Additionally to the single layer SQUIDs also flip-chip magnetometers were investigated. These sensors consist of a small washer-shaped read-out SQUID flipped to a coupling coil incorporated in a large (10 mm x 10 mm) pickup loop. Manufacturing and properties are described elsewhere [21], [22]. All measurements were performed in liquid nitrogen at 77 K.
Maximum voltage modulation $\Delta V_{MAX}$ and corresponding bias current, $I_{MAX}$, are measured directly; inductance, L, is determined from geometry by a numerical method described in [23], which also includes the kinetic inductance; the resistance, R, is determined from VCC of the DC SQUID in zero magnetic flux.

Since the critical current, $I_C$, of a high $T_C$ DC SQUID cannot be measured with a good accuracy due to high level of thermal fluctuations, we here, as distinct from [7], don't use it for normalization of the experimental voltage values. For the comparison with the theory we use only the quantities, which are measured directly, $\Delta V_{MAX}$ and $I_{MAX}$.

7The comparison of the theory with experiment is shown in Fig. 5. As is seen from the figure, most of experimental points lie well below the curve for DC SQUID with conventional CPR, however there exist few experimental points which lie higher the conventional CPR curve. As it follows from these graphs the theory can explain the large deviation of the experimental voltage modulation from theoretical curve for DC SQUID with conventional CPR. The theory cannot explain the points, which lie above conventional CPR curve, however, we hope they can be explained if the asymmetry in the amplitudes of the second harmonic $I_2^{(1)} \neq I_2^{(2)}$ is taken into account.

## 5 Conclusion

We have shown in the paper that the significant reduction of the voltage modulation in high $T_C$ DC SQUIDs, which lie well below the theoretical predictions and computer simulations based on conventional sinusoidal CPR, can be explained by relatively large amplitude of a second harmonic in the junction CPR.

Although, the existence of second harmonic in high $T_C$ grain boundary junctions follows from the theory [8, 9, 10], and there exist some mechanisms (phase fluctuations between d- and s-wave pairing; faceted structure of grain boundary) that can enhance the contribution of second harmonic [24, 25], but up till now a reliable evidence of substantial second harmonic was observed only in [001]-tilt grain boundary junction with 45º and 30° asymmetric misorientation angle [12, 13, 14, 15, 16]. Therefore, we cannot rule out other mechanisms, which lead to the reduction of the voltage modulation in high $T_C$ DC SQUIDs.

**Acknowledgement**

The authors are grateful to Evgeni Il'ichev and Miroslav Grajcar for fruitful discussions.
The work is partly supported by INTAS Program of EU under grant 2001-0809.

**Figure captions**

Fig. 1. Voltage-flux curves for small inductance DC SQUID. (solid square)- $I_1$=10 µA, $I_2$=0 (2-nd harmonic is absent), $i=I/I_1$; (solid circle)- $I_1$=10 µA, $I_2$=10 µA, $i=I/I_1$=1.5; (cross)- $I_1$=0, $I_2$=10 µA (1-st harmonic is absent), $i=I/I_2$=1.5;

Fig. 2. Voltage-current curves for large inductance DC SQUID. $L=2L_F$, $\Phi_X$=0, $\gamma=I_2/I_1$; (solid square)- $I_1$=20 µA, $I_2$=0; (open circle)- $I_1$=20 µA, $I_2$=0.3 $I_1$; (open triangle)- $I_2$=20 µA, $I_1$=0.3 $I_2$;

Fig. 3. The reduced voltage modulation vs bias current curves in the absence of the second harmonic. (solid square)-$I_1$=10 µA; (solid circle)-$I_1$=20 µA; (solid triangle)-$I_1$=30 µA.

Fig. 4. The influence of the second harmonic on the voltage modulation. (solid square)-$I_1$=20 µA, $I_2$=0; (open triangle)- $I_1$=15 µA, $I_2$= −5 µA; (open circle)- $I_1$=15 µA, $I_2$= 5 µA; (open star) $I_1$=5 µA, $I_2$= 15 µA; (*)-$I_1$=5 µA, $I_2$= −15 µA.

Fig. 5. Dependence of maximum voltage modulation on the bias current. Comparison with experiment. (solid square)-$I_2$=0; (open circle)-$I_2/I_1$=−0.1; (open triangle)- $I_2/I_1$=0.1; (open square)- $I_2/I_1$=0.3; (open star)-$I_2/I_1$=−0.3; (left corner open triangle)-$I_2/I_1$=0.5; (right corner open triangle)-$I_2/I_1$= −0.5; (solid triangle)-$I_2/I_1$=10; (solid circle)- $I_2/I_1$=3.3; (solid star)- $I_2/I_1$=2; (cross)-experimental points.

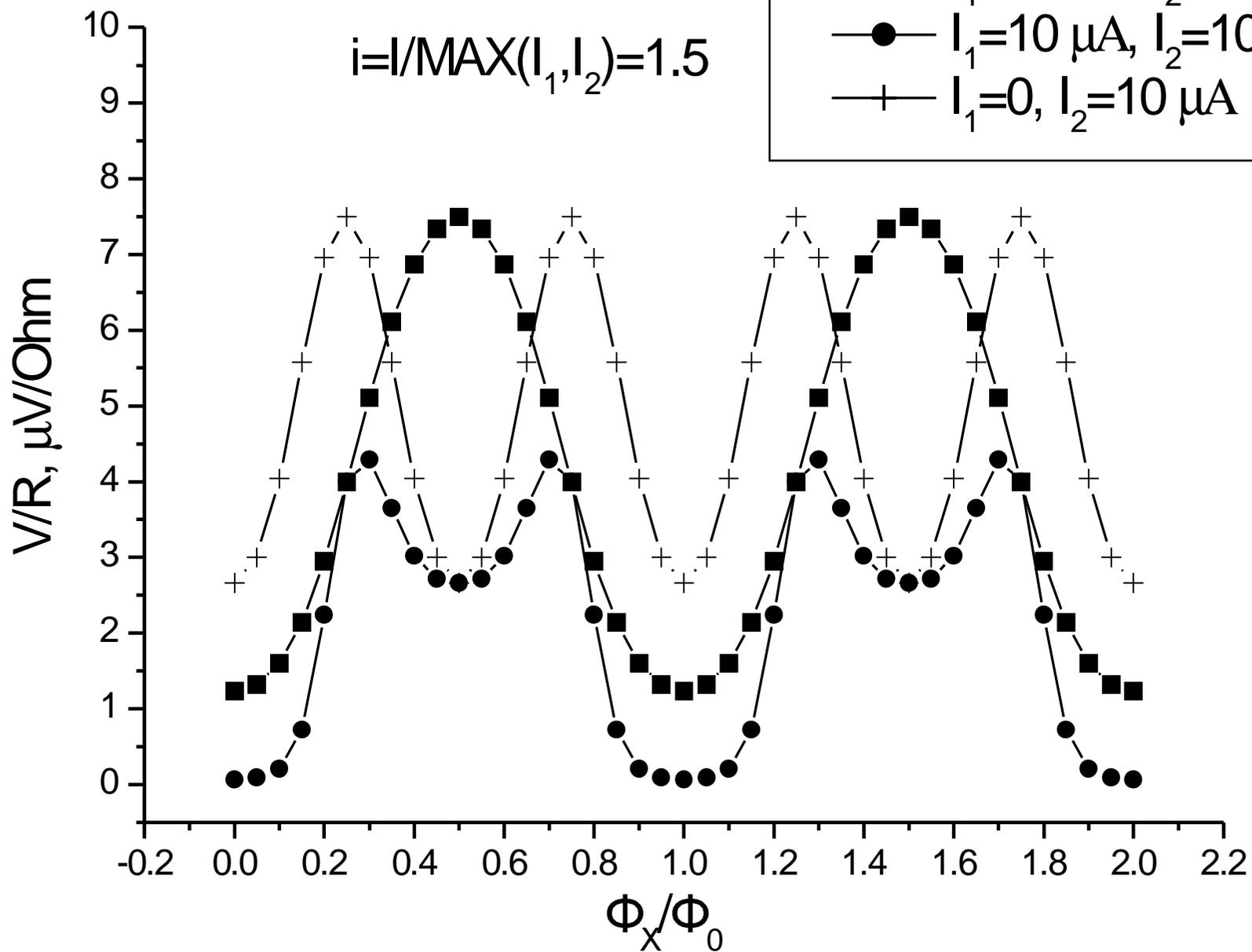
Fig. 1

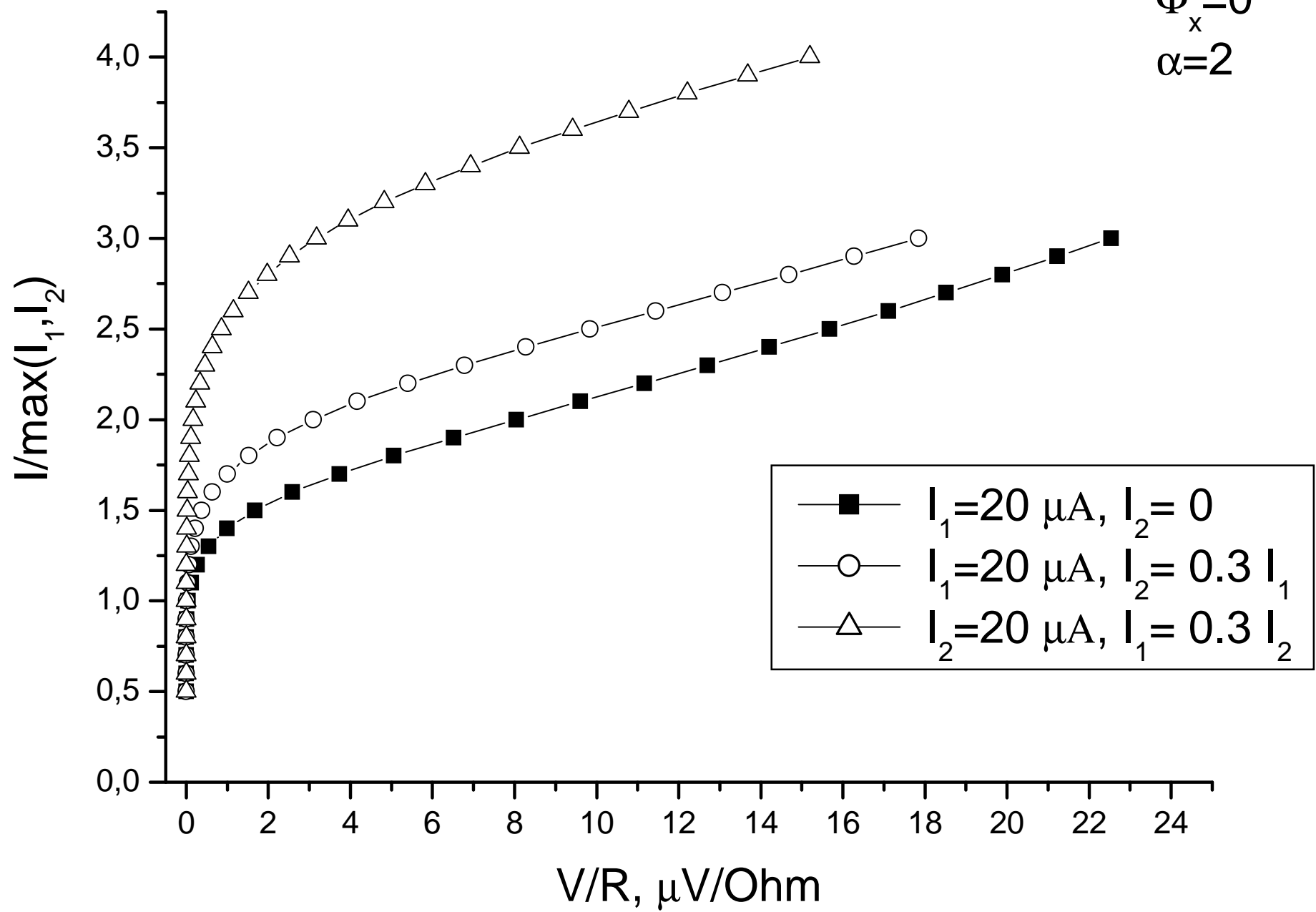

Fig. 2

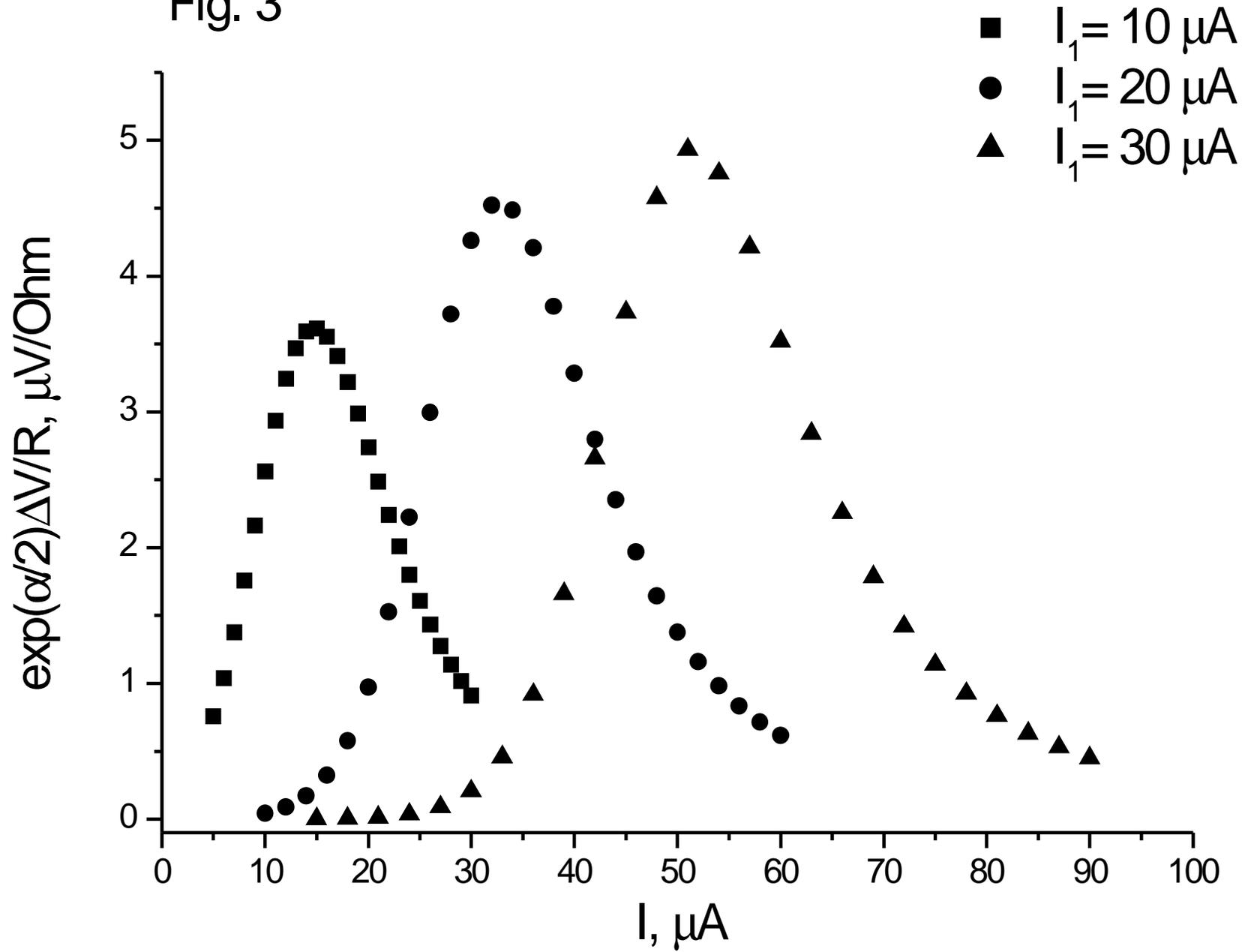

Fig. 3

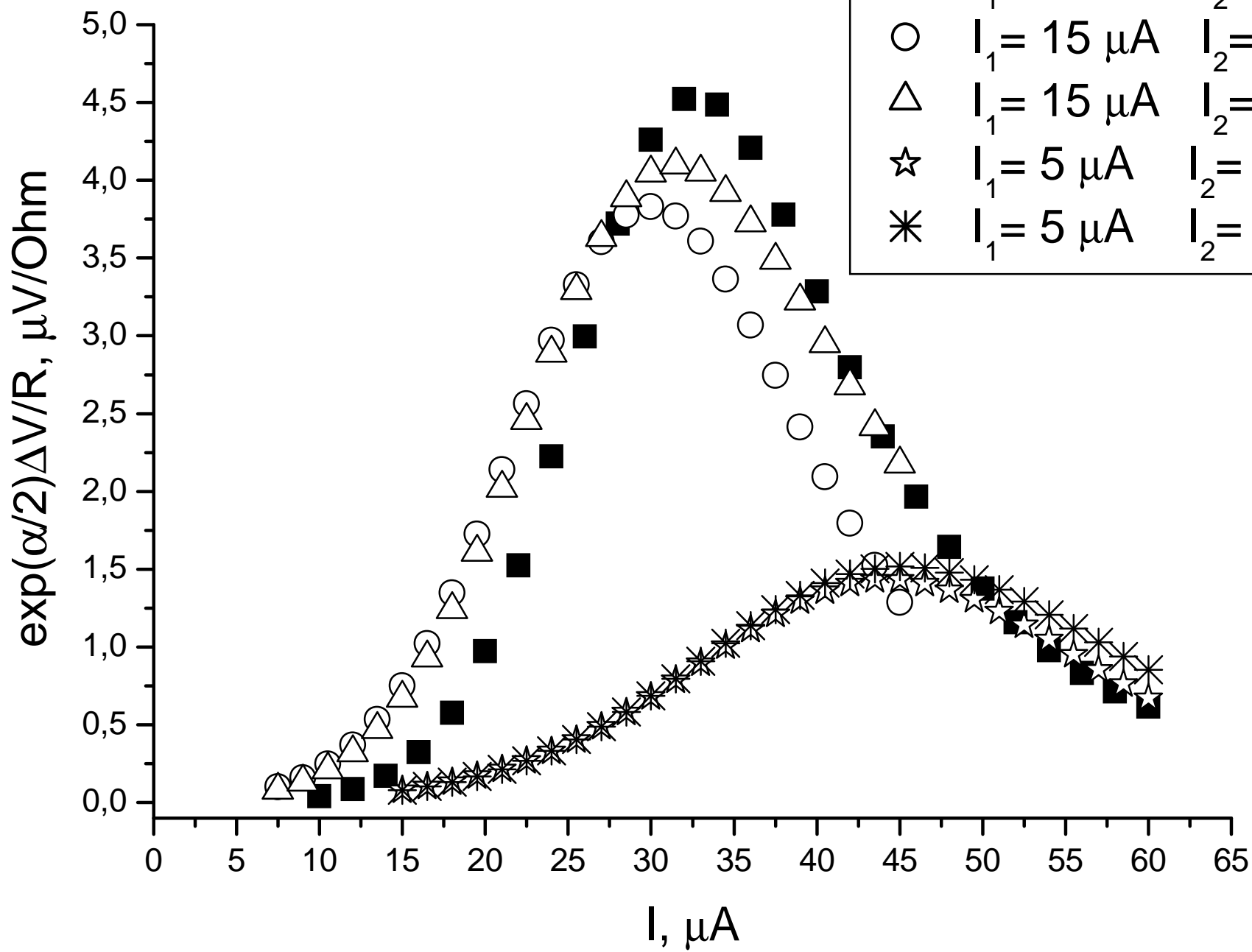

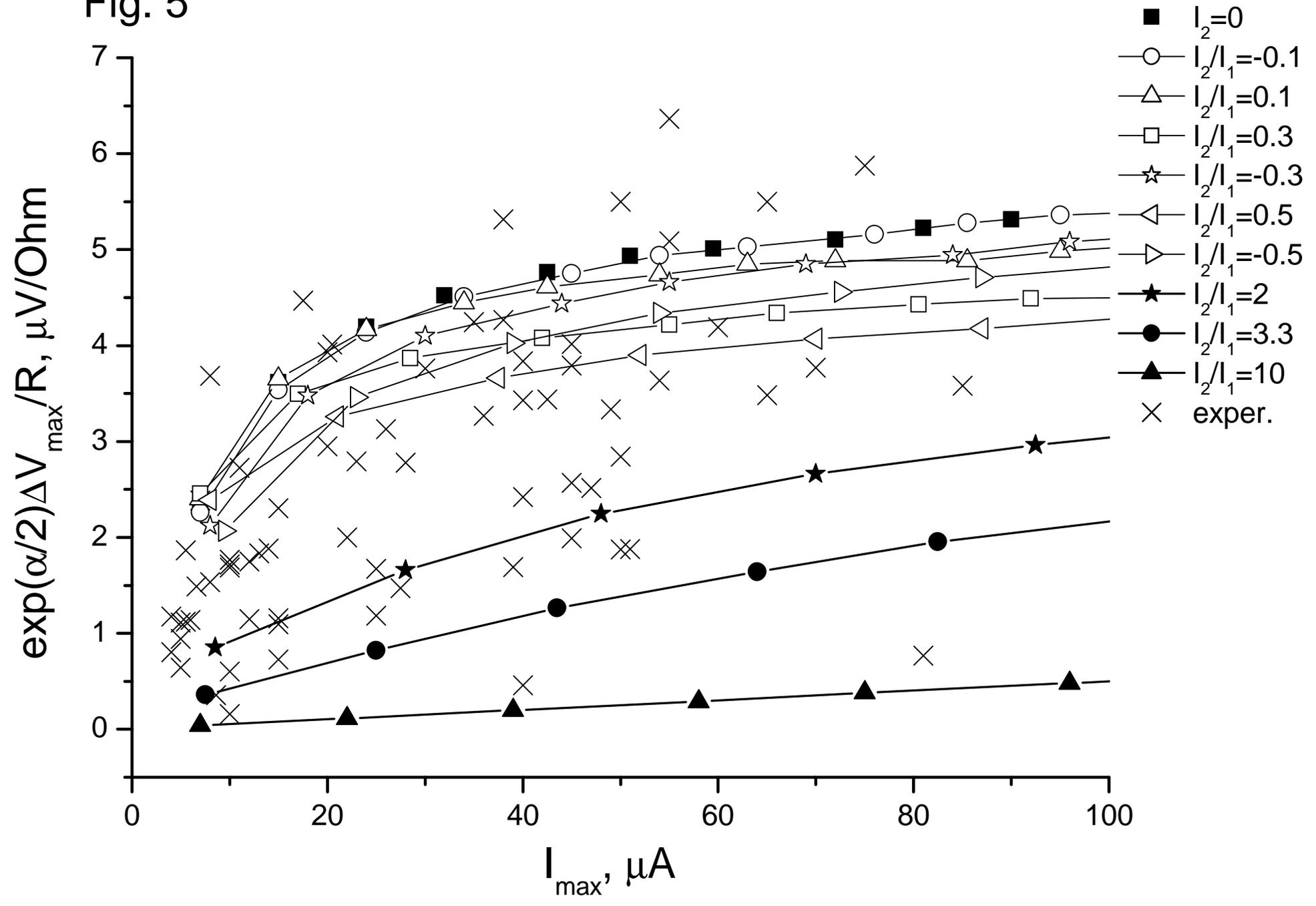

Fig. 5